\documentclass[aps,twocolumn,nofootinbib,showpacs,longbibliography]{revtex4-1}
\usepackage{bm,amsmath,amssymb,graphicx}
\newcommand{\Pbracket}[2]{\{#1,#2\}}
\newcommand{\qbarddd}{\mbox{$\rule{0mm}{2.0ex}^{\bm{\cdots}}\hspace{-0.75em}\bar{q}$}}

\begin{document}

\title{Hamiltonian formulation of a class of constrained fourth-order differential equations in the Ostrogradsky framework}

\author{Hans Christian \"Ottinger}
\email[]{hco@mat.ethz.ch}
\homepage[]{http://www.polyphys.mat.ethz.ch/}
\affiliation{ETH Z\"urich, Department of Materials, Polymer Physics, HCP F 47.2,
CH-8093 Z\"urich, Switzerland}

\date{\today}

\begin{abstract}
We consider a class of Lagrangians that depend not only on some configurational variables and their first time derivatives, but also on second time derivatives, thereby leading to fourth-order evolution equations. The proposed higher-order Lagrangians are obtained by expressing the variables of standard Lagrangians in terms of more basic variables and their time derivatives. The Hamiltonian formulation of the proposed class of models is obtained by means of the Ostrogradsky formalism. The structure of the Hamiltonians for this particular class of models is such that constraints can be introduced in a natural way, thus eliminating expected instabilities of the fourth-order evolution equations. Moreover, canonical quantization of the constrained equations can be achieved by means of Dirac's approach to generalized Hamiltonian dynamics.
\end{abstract}

\pacs{11.10.Ef}

\maketitle



\section{Introduction}
In a monumental article written in 1848 (in French), Mikhail Vasilyevich Ostrogradsky laid the foundations for the Lagrangian and Hamiltonian formulation of higher-order differential equations and pointed out their disposition to instability \cite{Ostrogradsky1850}. Modern applications of Lagrangians with higher derivatives in particle physics include investigations of possible deviations of electroweak vector-boson self-interactions from the standard model (see \cite{GrosseKnetter94} and references therein).

The present work on higher-order theories is motivated by attempts to introduce an alternative theory of gravity as a Yang-Mills theory based on the Lorentz group (see, for example the works \cite{Utiyama56,Yang74,hco231} spanning more than six decades, and references therein). The vector potentials of the Yang-Mills theory are no longer considered as primary fields, but rather as functions of a decomposition of the metric tensor including time derivatives (the vector potentials appear as a spin connection in the spirit of the Ashtekar variables proposed for a canonical approach to gravity \cite{Ashtekar86,Ashtekar87}). Therefore, it is natural to consider theories given by standard Lagrangians $L(q,\dot{q})$ where the variables $q(\bar{q},\dot{\bar{q}})$ and $\dot{q}(\bar{q},\dot{\bar{q}},\ddot{\bar{q}})$ are given in terms of more fundamental variables $\bar{q}$.

Based on this idea, we introduce a class of higher-order models in the Lagrangian (Section~\ref{secL}) and Hamiltonian (Section~\ref{secH}) settings, where the reformulation is achieved by means of the Ostrogradsky framework. The expected Ostrogradsky instability is cured by means of constraints which, for the proposed class of models, arise naturally (Section~\ref{secOI}).  We offer a number of concluding remarks (Section~\ref{secCR}), in particular, on the role of constraints and the canonical quantization of the proposed higher-order theories. In the appendices, the abstract general ideas are illustrated in the context of simple examples from mechanics (Appendix~\ref{appexample}) and field theory (Appendix~\ref{appexampleGR}).

\section{Lagrangian formulation}\label{secL}
We start our development from the first-order Lagrangian
\begin{equation}\label{Lagrangian}
   L(q,\dot{q}) = \frac{1}{2} m \dot{q}_i \dot{q}_i + \dot{q}_i \, u_i(q) - V(q) ,
\end{equation}
for the discrete set of variables $q_i$, $i=1, \ldots I$, where $q$ represents the list of all variables, $u_i(q)$, $V(q)$ are sufficiently smooth functions, and we make use of Einstein's summation convention (summation over indices occurring twice). This type of Lagrangian is not only very common for mechanical systems, but it covers also the space-discretized version of the Yang-Mills Lagrangian \cite{YangMills54}.

As a next step, we assume that the variables $q$ can be expressed in terms of the variables $\bar{q}$, which are typically fewer than the variables $q$. Moreover, the variables $q$ are allowed to depend also on the time derivatives of $\bar{q}$,
\begin{equation}\label{variabletransf}
   q_i = \alpha_i(\bar{q}) + \beta_{ik}(\bar{q}) \, \dot{\bar{q}}_k ,
\end{equation}
where $\alpha_i(\bar{q})$, $\beta_{ik} (\bar{q})$ are sufficiently smooth functions of the variables $\bar{q}_k$, $k=1, \ldots K \leq I$. For dynamic consistency reasons, we postulate
\begin{equation}\label{variabletransff}
   \dot{q}_i = \alpha'_{ik}(\bar{q}) \, \dot{\bar{q}}_k
   + \beta'_{ikl}(\bar{q}) \, \dot{\bar{q}}_k \dot{\bar{q}}_l
   + \beta_{ik}(\bar{q}) \, \ddot{\bar{q}}_k ,
\end{equation}
with the derivatives
\begin{equation}\label{albeder}
   \alpha'_{ik}(\bar{q}) = \frac{\partial \alpha_i(\bar{q})}{\partial \bar{q}_k} , \qquad
   \beta'_{ikl}(\bar{q}) = \frac{\partial \beta_{ik}(\bar{q})}{\partial \bar{q}_l} .
\end{equation}
In the following, we use an analogous notation for the second-order derivatives of $\alpha_i(\bar{q})$ and $\beta_{ik}(\bar{q})$. According to Eqs.~(\ref{variabletransf}) and (\ref{variabletransff}), the Lagrangian (\ref{Lagrangian}) can be considered as a function of $\bar{q}$, $\dot{\bar{q}}$ and $\ddot{\bar{q}}$.

Note that the variables $q$, $\dot{q}$ represent $2I$ degrees of freedom, whereas the variables $\bar{q}$, $\dot{\bar{q}}$, $\ddot{\bar{q}}$ represent $3K$ degrees of freedom. A particularly interesting situation arises for $I=(3/2)K$. We can then consider the case where there is a one-to-one correspondence between the variables $q$, $\dot{q}$ and $\bar{q}$, $\dot{\bar{q}}$, $\ddot{\bar{q}}$. In other words, we can assume that the functions (\ref{variabletransf}), (\ref{variabletransff}) can be inverted to obtain $\bar{q}$, $\dot{\bar{q}}$, $\ddot{\bar{q}}$ uniquely from $q$, $\dot{q}$. An example of such an invertible relationship is given in Appendix~\ref{appexample} (all the steps of the general development are illustrated for that example in the appendix). In general, we make the regularity assumption that the rank of the matrix $\beta_{ik}$ in Eq.~(\ref{variabletransf}) takes its maximum possible value, $K$.

Stationarity of the time integral of the Lagrangian (\ref{Lagrangian}), or action, with respect to variations of $q$ leads to the $I$ evolution equations
\begin{equation}\label{Lagrangeqevol}
   m \ddot{q}_i + \omega_{ij} \dot{q}_j + \frac{\partial V}{\partial q_i} = 0 ,
\end{equation}
with
\begin{equation}\label{omegadef}
   \omega_{ij}(q) = \frac{\partial u_i(q)}{\partial q_j}
   - \frac{\partial u_j(q)}{\partial q_i}.
\end{equation}
If the variations of $q$ are restricted to the variations of $\bar{q}$ in Eqs.~(\ref{variabletransf}) and (\ref{variabletransff}), we obtain the following smaller set of $K$ evolution equations
\begin{equation}\label{Lagrangeqbarevol}
   \left ( \alpha'_{ik} + \beta'_{ilk} \, \dot{\bar{q}}_l
   - \frac{d}{dt} \beta_{ik} \right )
   \left ( m \ddot{q}_i + \omega_{ij} \dot{q}_j + \frac{\partial V}{\partial q_i} \right ) = 0 .
\end{equation}
Note that Eq.~(\ref{Lagrangeqbarevol}) contains third-order time derivatives of $q$, implying a set of fourth-order differential equations for $\bar{q}$. This is the class of fourth-order differential equations considered in this paper. As a consequence of the chain rule, they have the factorized structure of Eq.~(\ref{Lagrangeqbarevol}) because they result from second-order differential equations by considering the unknowns as functions of potentially fewer, more basic variables and their time derivatives. Our further investigation is motivated by the question whether there is a canonical quantization procedure for this class of fourth-order equations.

One would like solutions $q(t)$ of the second-order equations (\ref{Lagrangeqevol}) to provide solutions $\bar{q}(t)$ of the fourth-order equations (\ref{Lagrangeqbarevol}). Even if we assume that one can uniquely reconstruct $\bar{q}$, $\dot{\bar{q}}$, $\ddot{\bar{q}}$ from $q$, $\dot{q}$, this is not straightforward because the resulting $\dot{\bar{q}}(t)$, $\ddot{\bar{q}}(t)$ must be consistent with the time derivatives of $\bar{q}(t)$. Consistency is most easily achieved for static solutions. In general, symmetries are required to obtain valid solutions $\bar{q}(t)$ (see the example of Appendix~\ref{appexample}). Note that Eq.~(\ref{Lagrangeqevol}) is a system of third-order differential equation for $\bar{q}(t)$ consisting of $I$ equations for $K \leq I$ functions.

\section{Hamiltonian formulation}\label{secH}
We next consider the Hamiltonian formulation of the fourth-order differential equations (\ref{Lagrangeqbarevol}) for $\bar{q}(t)$. Such a formulation can be achieved by means of the Ostrogradsky framework. The key idea is to use
\begin{equation}\label{ostrovarQ12}
   Q_{1 k} = \bar{q}_k , \quad Q_{2 k} = \dot{\bar{q}}_k ,
\end{equation}
as configurational variables and to define the corresponding conjugate momenta by
\begin{eqnarray}
   P_{1 k} = \frac{\partial L}{\partial \dot{\bar{q}}_k}
   - \frac{d}{dt} \frac{\partial L}{\partial \ddot{\bar{q}}_k} &=&
   - \left ( m \ddot{q}_i + \omega_{ij} \dot{q}_j
   + \frac{\partial V}{\partial q_i} \right ) \beta_{ik} \qquad
   \nonumber \\ &+&
   (m \dot{q}_i+u_i) ( \alpha'_{ik} + \beta'_{ilk} \, \dot{\bar{q}}_l ) ,
\label{ostrovarP1}
\end{eqnarray}
and
\begin{equation}\label{ostrovarP2}
   P_{2 k} = \frac{\partial L}{\partial \ddot{\bar{q}}_k} = (m \dot{q}_i+u_i) \beta_{ik} .
\end{equation}
Note that $P_{2 k}$ contains second-order time derivatives of $\bar{q}_k$, whereas $P_{1 k}$ contains third-order time derivatives of $\bar{q}_k$.

In view of Eqs.~(\ref{variabletransf}), (\ref{variabletransff}) and (\ref{ostrovarQ12}), the key step in calculating the Hamiltonian $H = P_{1 k} \dot{Q}_{1 k} + P_{2 k} \dot{Q}_{2 k} - L$ by Legendre transformation is the determination of $\ddot{\bar{q}}_k(Q_1,Q_2,P_2)$. Equation~(\ref{ostrovarP2}), combined with Eq.~(\ref{variabletransff}), gives
\begin{equation}\label{qbarddotsource}
   \beta_{ik} \beta_{il} \, \ddot{\bar{q}}_l = \frac{1}{m} ( P_{2 k} - u_i \beta_{ik} )
   - (\alpha'_{in} + \beta'_{iln} \, \dot{\bar{q}}_l) \dot{\bar{q}}_n \beta_{ik} .
\end{equation}
According to our regularity assumption, the symmetric $K \times K$ matrix $\beta_{ik} \beta_{il}$ has an inverse $B_{kl}$. We hence find the desired relation for $\ddot{\bar{q}}_k(Q_1,Q_2,P_2)$,
\begin{equation}\label{qbarddot}
   \ddot{\bar{q}}_k = \left\{ \frac{1}{m} P_{2 m}
   - \Big[ \frac{u_i}{m} + (\alpha'_{in} + \beta'_{iln} \,
   \dot{\bar{q}}_l) \dot{\bar{q}}_n \Big] \beta_{im} \right\} B_{mk} ,
\end{equation}
where $\dot{\bar{q}}_l=Q_{2 l}$, the functions $\beta_{im}$, $\alpha'_{in}$, $\beta'_{iln}$, $B_{mk}$ depend on $Q_1$ and, finally, $u_i$ is a function of $q_i(Q_1,Q_2)$. In a similar way, we find the third-order time derivatives
\begin{eqnarray}
   \qbarddd_k &=&  - \bigg \{ \frac{1}{m} P_{1 m}
   - \Big( \dot{q}_i + \frac{u_i}{m} \Big) ( \alpha'_{im} + \beta'_{ilm} \, \dot{\bar{q}}_l )
   \nonumber \\
   &+& \bigg [ \frac{1}{m} \Big ( \omega_{ij} \dot{q}_j + \frac{\partial V}{\partial q_i} \Big )
   + (\alpha''_{iln} + \beta''_{il\tilde{n}n} \dot{\bar{q}}_{\tilde{n}}) \,
   \dot{\bar{q}}_l \dot{\bar{q}}_n
   \nonumber \\
   &+& (\alpha'_{in} + \beta'_{iln} \dot{\bar{q}}_l + 2 \beta'_{inl} \dot{\bar{q}}_l)
   \, \ddot{\bar{q}}_n \bigg ] \beta_{im} \bigg \} B_{mk} .
\label{qbardddot}
\end{eqnarray}
At this point, with Eqs.~(\ref{ostrovarQ12}), (\ref{qbarddot}) and (\ref{qbardddot}), we have fully established the one-to-one relation between the variables $\bar{q}$, $\dot{\bar{q}}$, $\ddot{\bar{q}}$ and $\qbarddd$ and the canonical variables $Q_1$, $Q_2$, $P_1$ and $P_2$. In particular, we can also find $q$, $\dot{q}$ and $\ddot{q}$ in terms of the canonical variables.

The Hamiltonian now takes the following form,
\begin{eqnarray}
   H &=& \frac{1}{2} m \dot{q}_i \dot{q}_i + V(q)
   \nonumber \\
   &+& \Big[ P_{1 k} - (m \dot{q}_i+u_i) (\alpha'_{ik}
   + \beta'_{ilk} \, Q_{2 l}) \Big] Q_{2 k} . \qquad
\label{Hamiltonian}
\end{eqnarray}
In order to find the derivatives of $H$ with respect to the canonical variables we need to use the substitution rules for $q$ and $\dot{q}$. The following auxiliary results are helpful:
\begin{equation}\label{deraux1}
   \beta_{ik} \, d ( m \dot{q}_i + u_i ) = d P_{2 k} - (m \dot{q}_i+u_i) \beta'_{ikl} \, d Q_{1 l} ,
\end{equation}
and
\begin{eqnarray}
   d H &=& P_{1 k} d Q_{2 k} + Q_{2 k} d P_{1 k} + \left ( \frac{\partial V}{\partial q_i}
   -  \frac{\partial u_j}{\partial q_i} \dot{q}_j \right ) d q_i
   \nonumber \\
   &+& \ddot{\bar{q}}_k \beta_{ik} \, d ( m \dot{q}_i + u_i )
   - (m \dot{q}_i+u_i) \, d ( \alpha'_{ik} + \beta'_{ilk} \, \dot{\bar{q}}_l ) \dot{\bar{q}}_k .
   \nonumber \\ &&
\label{deraux2}
\end{eqnarray}
In deriving these auxiliary equations, only Eq.~(\ref{qbarddotsource}) and not Eq.~(\ref{qbarddot}) has been used, so that the following results for the derivatives of $H$ hold even without regularity assumption,
\begin{equation}\label{dHamiltoniandP1}
   \frac{\partial H}{\partial P_{1 k}} = \dot{\bar{q}}_k ,
\end{equation}
\begin{equation}\label{dHamiltoniandP2}
   \frac{\partial H}{\partial P_{2 k}} = \ddot{\bar{q}}_k ,
\end{equation}
\begin{eqnarray}
   \frac{\partial H}{\partial Q_{1 k}} &=& \left ( \frac{\partial V}{\partial q_i}
   -  \frac{\partial u_j}{\partial q_i} \dot{q}_j \right )
   ( \alpha'_{ik} + \beta'_{ilk} \, \dot{\bar{q}}_l )
   \nonumber \\
   &-& (m \dot{q}_i+u_i) \, \frac{d}{d t} ( \alpha'_{ik} + \beta'_{ilk} \, \dot{\bar{q}}_l ) ,
\label{dHamiltoniandQ1}
\end{eqnarray}
\begin{eqnarray}
   \frac{\partial H}{\partial Q_{2 k}} &=& P_{1 k} + \left ( \frac{\partial V}{\partial q_i}
   -  \frac{\partial u_j}{\partial q_i} \dot{q}_j \right ) \beta_{ik}
   \nonumber \\
   &-& (m \dot{q}_i+u_i) ( \alpha'_{ik} + \beta'_{ilk} \, \dot{\bar{q}}_l
   + \beta'_{ikl} \, \dot{\bar{q}}_l ) . \qquad
\label{dHamiltoniandQ2}
\end{eqnarray}
Some of the terms in Eqs.~(\ref{dHamiltoniandQ1}) and (\ref{dHamiltoniandQ2}) can be written in a simpler form, directly in terms of canonical variables. For example, if we introduce $\bar{V}(\bar{q}, \dot{\bar{q}}) = V(q(\bar{q},\dot{\bar{q}}))$, then we can write
\begin{equation}\label{genVsimpl1}
   \frac{\partial V(q)}{\partial q_i}
   ( \alpha'_{ik} + \beta'_{ilk} \, \dot{\bar{q}}_l ) =
   \frac{\partial \bar{V}(Q_1,Q_2)}{\partial Q_1} ,
\end{equation}
and
\begin{equation}\label{genVsimpl2}
   \frac{\partial V(q)}{\partial q_i} \beta_{ik} =
   \frac{\partial \bar{V}(Q_1,Q_2)}{\partial Q_2} .
\end{equation}
The same type of simplifications can also be achieved for $\partial u_j/\partial q_i$. On the other hand, $\dot{q}_i$ is a fairly complicated function of the canonical variables, which involves the lengthy expression (\ref{qbarddot}) for $\ddot{\bar{q}}_k$, to be inserted into Eq.~(\ref{variabletransff}). In view of this complexity, the auxiliary result (\ref{deraux1}) is remarkably simple.

The canonical equations $\dot{Q}_1 = \partial H/\partial P_1$ and $\dot{Q}_2 = \partial H/\partial P_2$ are consistent with the definitions (\ref{ostrovarQ12}) of $Q_1$ and $Q_2$. As $\dot{P}_2 = -\partial H/\partial Q_2$ reproduces the representation (\ref{ostrovarP1}) of $P_1$, the evolution equation must be contained in $\dot{P}_1 = -\partial H/\partial Q_1$ (note that $\dot{P}_1$ contains third-order time derivatives of $q$ and hence fourth-order derivatives of $\bar{q}$). Indeed, by using Eqs.~(\ref{dHamiltoniandQ1}) and (\ref{ostrovarP1}), we recover the fourth-order evolution equation (\ref{Lagrangeqbarevol}).

\section{Ostrogradsky instability}\label{secOI}
The occurrence of the term $P_{1 k} Q_{2 k}$ in the Hamiltonian (\ref{Hamiltonian}) implies that the energy can be lowered without any bound by increasing the momentum $P_{1 k}$ to large positive or negative values. Therefore, this term suggests an instability that is known as Ostrogradsky instability. Such an instability is generally considered as a strong argument against higher-order equations.

The unbounded term in the Hamiltonian can be suppressed by imposing the $K$ primary constraints
\begin{equation}\label{genconstraints1}
   \varphi^{(1)}_k = P_{1 k} - (m \dot{q}_i+u_i) (\alpha'_{ik}
   + \beta'_{ilk} \, \dot{\bar{q}}_l) = 0 .
\end{equation}
The potentially nice features of these constraints can be recognized by a closer look at Eq.~(\ref{ostrovarP1}). This expression for $P_{1 k}$ implies that the primary constraints (\ref{genconstraints1}) are fulfilled by the solutions of the basic equation (\ref{Lagrangeqevol}), which we want to keep, but not necessarily by the solutions of the higher-order equation (\ref{Lagrangeqbarevol}), from which we want to eliminate instabilities. On a more formal level, if the primary constraints (\ref{genconstraints1}) are imposed, the constrained Hamiltonian (\ref{Hamiltonian}) is obtained by the same substitution idea as the Lagrangian: the variables $q$, $\dot{q}$ are given in terms of $\bar{q}$, $\dot{\bar{q}}$, $\ddot{\bar{q}}$ in Eqs.~(\ref{variabletransf}), (\ref{variabletransff}) and, in turn, $\bar{q}$, $\dot{\bar{q}}$, $\ddot{\bar{q}}$ are expressed in terms of the canonical variables $Q_1$, $Q_2$, $P_2$ according to Eqs.~(\ref{ostrovarQ12}) and (\ref{qbarddot}). The constrained Hamiltonian does not depend on $P_1$ but, according to Eq.~(\ref{dHamiltoniandQ2}), the derivatives of the Hamiltonian on the constrained manifold do.

The primary constraints are not consistent with the dynamics,
\begin{equation}\label{genconstraintsder}
   \dot{\varphi}^{(1)}_k = - \left ( \alpha'_{ik} + \beta'_{ilk} \, \dot{\bar{q}}_l \right )
   \left ( m \ddot{q}_i + \omega_{ij} \dot{q}_j + \frac{\partial V}{\partial q_i} \right ) ,
\end{equation}
so that we introduce the secondary constraints
\begin{equation}\label{genconstraints2}
   \varphi^{(2)}_k = \dot{\varphi}^{(1)}_k = 0 .
\end{equation}
These secondary constraints share the potentially nice features of the primary constraints, keeping physical solutions but not all terms of the fourth-order equations.

The explicit example of Appendix~\ref{appexample} shows that it may be necessary to continue the iterative procedure and to consider also the tertiary constraints
\begin{equation}\label{genconstraints3}
   \varphi^{(3)}_k = \dot{\varphi}^{(3)}_k = 0 .
\end{equation}
In general, the iterative procedure needs to be continued until full dynamic consistency is reached on the constrained manifold.

\section{Summary and conclusions}\label{secCR}
We have introduced a class of Lagrangians $L(\bar{q},\dot{\bar{q}},\ddot{\bar{q}})$ associated with fourth-order evolution equations by substituting a transformation $q(\bar{q},\dot{\bar{q}})$, as well as the consistent transformation $\dot{q}(\bar{q},\dot{\bar{q}},\ddot{\bar{q}})$, into a standard Lagrangian $L(q,\dot{q})$. The corresponding canonical Hamiltonian formulation with Hamiltonian $H(Q_1,Q_2,P_1,P_2)$ is obtained by means of the Ostrogradsky formalism. Natural constraints arise from the idea that, on the constrained manifold, the Hamiltonian of the higher-order problem should be obtained from a substitution procedure, just like the Lagrangian.

These natural constraints play a crucial role in the proposed class of models. They are needed to eliminate the Ostrogradsky instability that one expects because the Hamiltonian contains the momentum $P_1$ only in a linear term and hence is unbounded. Ideally, the constraints restrict the solutions of the higher-order problem to a subset of solutions of the original standard problem. This requirement is not fulfilled automatically and imposes restrictions on the original Lagrangian and its interplay with the transformation; the transformation should be consistent with the symmetries of the Lagrangian. If the higher-order formulation only picks out solutions from the standard formulation, the most interesting features may arise only upon coupling to other systems.

In general, the constraints have nonvanishing Poisson brackets so that they may be classified as second-class constraints. For second-class constraints, the Poisson bracket can be modified into a Dirac bracket that leads to a canonical quantization procedure \cite{Dirac50,Dirac58a,Dirac58b}. Compared to an alternative quantization scheme based on reducing general higher-order Lagrangians to first-order Lagrangians proposed in \cite{Nawafleh11}, we here exploit the special features of our particular class of higher-order Lagrangians.

There may be further constraints already present in the original theory represented by $L(q,\dot{q})$, for example, for gauge theories. The gauge transformation of the $q$ variables should then be inherited from the gauge transformation behavior of the $\bar{q}$ variables. In the quantization procedure, the constraints associated with gauge transformations can be handled by the powerful BRST procedure \cite{Nemeschanskyetal86,PeskinSchroeder,hco229}, where the acronym BRST refers to Becchi, Rouet, Stora \cite{BecchiRouetStora76} and Tyutin \cite{Tyutin75}.

There is an interesting epistemological aspect of the present work. The class of higher-order models proposed in this paper is particularly promising in situations where a successful theory should be modified because the fundamental objects (``particles'') of the theory are discovered to be composed of even more fundamental objects. For example, the idea to decompose the space-time metric $g_{\mu\nu}$ of general relativity as
$g_{\mu\nu} = \eta_{\kappa\lambda} {b^\kappa}_\mu {b^\lambda}_\nu$
in terms of the Minkowski metric $\eta_{\kappa\lambda}$ and the more fundamental variables ${b^\kappa}_\mu$ has profound implications for the theory of gravity \cite{Ashtekar86,Ashtekar87}. In particular, the variables $A_{a\nu}$ characterizing the connection of the variables ${b^\kappa}_\mu$ (``spin-connection,'' similar to the Levi-Civita connection for $g_{\mu\nu}$) can be considered as a most relevant example of a mapping $q(\bar{q},\dot{\bar{q}})$ with $q=A_{a\nu}$ and $\bar{q}={b^\kappa}_\mu$, where the subscript $a$ labels the generators of a Lie group (a simplified version of this field theoretic example is presented in Appendix~\ref{appexampleGR}). If the Lie group is given by the Lorentz group \cite{hco231}, the index $a$ takes six values, compared to the four values of the space-time index $\kappa$; we thus find the ratio $3:2$ for the natural Lorentz-group/space-time pairing, which has been revealed to be of special relevance in one-to-one reformulations of standard Lagrangian theories. In terms of the variables ${b^\kappa}_\mu$, one can formulate a dissipative quantum field theory based on a dynamic ``diffusive smearing'' mechanism on the Planck scale \cite{hcoqft}.

\appendix

\section{Example from mechanics} \label{appexample}
For the linear transformation
\begin{equation}\label{exlintransf}
   \left(
     \begin{array}{c}
       q_1 \\
       q_2 \\
       q_3 \\
       \dot{q}_1 \\
       \dot{q}_2 \\
       \dot{q}_3 \\
     \end{array}
   \right) =
   \left(
     \begin{array}{llllll}
       1 & 0    & 0 & \lambda    &   &   \\
       0 & 1    & 0 & 0    &   &   \\
       0 & 0 \; & \lambda & 0    &   &   \\
         &      & 1 & 0    & 0 & \lambda \\
         &      & 0 & 1    & 0 & 0 \\
         &      & 0 & 0 \; & \lambda & 0 \\
     \end{array}
   \right)
   \left(
     \begin{array}{c}
       \bar{q}_1 \\
       \bar{q}_2 \\
       \dot{\bar{q}}_1 \\
       \dot{\bar{q}}_2 \\
       \ddot{\bar{q}}_1 \\
       \ddot{\bar{q}}_2 \\
     \end{array}
   \right) ,
\end{equation}
one can easily verify that it is invertible for $\lambda \neq 0$, where $\lambda$ has the dimensions of a time constant. Moreover, the consistency condition (\ref{variabletransff}) is satisfied. For the three-dimensional harmonic oscillator with mass $m$ and spring constants $h_i$ with
\begin{equation}\label{exLstandard}
   L = \frac{1}{2} m \left( \dot{q}_1^2 + \dot{q}_2^2 + \dot{q}_3^2 \right)
   - \frac{1}{2} \left( h_1 q_1^2 + h_2 q_2^2 + h_3 q_3^2 \right) ,
\end{equation}
which is of the general form (\ref{Lagrangian}), the higher-order Lagrangian obtained by insertion becomes
\begin{eqnarray}
   L &=& \frac{1}{2} m \left[ \dot{\bar{q}}_1^2 + \dot{\bar{q}}_2^2
   + \lambda^2(\ddot{\bar{q}}_1^2 + \ddot{\bar{q}}_2^2) + 2 \lambda \dot{\bar{q}}_1 \ddot{\bar{q}}_2 \right]
   \nonumber \\
   &-& \frac{1}{2} \left[ h_1 \bar{q}_1^2 + h_2 \bar{q}_2^2 + \lambda^2(h_3 \dot{\bar{q}}_1^2
   + h_1 \dot{\bar{q}}_2^2) + 2 \lambda h_1 \bar{q}_1 \dot{\bar{q}}_2 \right] .
   \nonumber \\ &&
\label{exLhigher}
\end{eqnarray}
Instead of the usual three second-order equations (no summation over $i$),
\begin{equation}\label{exLaguseq}
   m \ddot{q}_i + h_i q_i = 0 ,
\end{equation}
we find the two fourth-order equations for $\bar{q}_1$ and $\bar{q}_2$,
\begin{eqnarray}
   h_1 ( \bar{q}_1 + \lambda \dot{\bar{q}}_2 )
   &+& \frac{d}{d t} \Big[ m ( \dot{\bar{q}}_1 + \lambda \ddot{\bar{q}}_2 )
   - \lambda^2 h_3 \dot{\bar{q}}_1 \Big]
   \nonumber \\
   &-& \lambda^2 \frac{d^2}{d t^2} m \ddot{\bar{q}}_1 = 0 ,
   \nonumber \\
   h_2 \bar{q}_2
   &+& \frac{d}{d t} \Big[ m \dot{\bar{q}}_2 - \lambda h_1 ( \bar{q}_1 + \lambda \dot{\bar{q}}_2 ) \Big]
   \nonumber \\
   &-& \lambda \frac{d^2}{d t^2} m ( \dot{\bar{q}}_1 + \lambda \ddot{\bar{q}}_2 ) = 0 ,
\label{exLageq4}
\end{eqnarray}
which can be rewritten in the form (\ref{Lagrangeqbarevol}),
\begin{eqnarray}
   m \ddot{q}_1 + h_1 q_1
   - \lambda \frac{d}{d t} ( m \ddot{q}_3 + h_3 q_3 ) &=& 0 ,
   \nonumber \\
   m \ddot{q}_2 + h_2 q_2
   - \lambda \frac{d}{d t} ( m \ddot{q}_1 + h_1 q_1 ) &=& 0 .
\label{exLageq3}
\end{eqnarray}

Solution of the usual equations (\ref{exLaguseq}) for three harmonic oscillators requires six initial conditions. The solutions are of the form
\begin{equation}\label{solqi}
   q_i = c_i \cos \omega_i t + c'_i \sin \omega_i t ,
\end{equation}
where $\omega_i=\sqrt{h_i/m}$ for $i=1,2,3$. Using to the inverse of the transformation (\ref{exlintransf}), these solutions suggest
\begin{eqnarray}
   \bar{q}_1 \!\! &=& \!\! c_1 \cos \omega_1 t + c'_1 \sin \omega_1 t
   + \lambda\omega_2 ( c_2 \sin \omega_2 t - c'_2 \cos \omega_2 t ) ,
   \nonumber \\
   \bar{q}_2 \!\! &=& \!\! c_2 \cos \omega_2 t + c'_2 \sin \omega_2 t ,
\label{solqbari}
\end{eqnarray}
as candidates for solving the higher-order equations. In general, however, these functions $\bar{q}_1$, $\bar{q}_2$ do not provide a solution to Eq.~(\ref{exLageq4}). Only for equal $h_i$ ($h_i=h$, $\omega=\sqrt{h/m}$), the functions given in Eq.~(\ref{solqbari}) provide a four-parameter solution to Eq.~(\ref{exLageq4}). Note that the transformation (\ref{exlintransf}) breaks the symmetry between $q_1$, $q_2$ and $q_3$ in an unnatural way so that, in the anisotropic case, the higher-order problem becomes completely different from the lower-order problem. From now on, we therefore restrict ourselves to the isotropic case.

On the one hand, for $\omega_1=\omega_2$, the solution (\ref{solqbari}) for $\bar{q}_1$, $\bar{q}_2$ represents only two independent harmonic oscillators with four parameters, whereas the underlying $q_1$, $q_2$, $q_3$ represent three independent harmonic oscillators with six parameters. On the other hand, we would expect eight parameters in the solution of the two fourth-order equations (\ref{exLageq4}), which we indeed recognize in the general solution of the system in Eq.~(\ref{exLageq4}) for the isotropic case,
\begin{eqnarray}
   \bar{q}_1 &=& \bar{c}_1 \cos \omega t + \bar{c}_2 \sin \omega t \nonumber \\
   &+& e^{-\sqrt{3}t/(2\lambda)}
   \left( \bar{c}_5 \cos \frac{t}{2\lambda} + \bar{c}_6 \sin \frac{t}{2\lambda} \right)
   \nonumber \\
   &+& e^{\sqrt{3}t/(2\lambda)}
   \left( \bar{c}_7 \cos \frac{t}{2\lambda} + \bar{c}_8 \sin \frac{t}{2\lambda} \right) ,
\label{solqbarigen1}
\end{eqnarray}
and
\begin{eqnarray}
   \bar{q}_2 &=& \bar{c}_3 \cos \omega t + \bar{c}_4 \sin \omega t \nonumber \\
   &+& e^{-\sqrt{3}t/(2\lambda)}
   \left( \bar{c}_6 \cos \frac{t}{2\lambda} - \bar{c}_5 \sin \frac{t}{2\lambda} \right)
   \nonumber \\
   &+& e^{\sqrt{3}t/(2\lambda)}
   \left( \bar{c}_8 \cos \frac{t}{2\lambda} - \bar{c}_7 \sin \frac{t}{2\lambda} \right) .
\label{solqbarigen2}
\end{eqnarray}
The existence of exponentially increasing and decreasing contributions with rates and frequencies independent of $m$ and $h$ is based on the two identities
\begin{equation}\label{exsolids}
   \bar{q}_1 + \lambda \dot{\bar{q}}_2 - \lambda^2 \ddot{\bar{q}}_1 = 0 , \qquad
   \bar{q}_2 - \lambda \dot{\bar{q}}_1 - \lambda^2 \ddot{\bar{q}}_2 = 0 ,
\end{equation}
which lead to solutions of Eq.~(\ref{exLageq4}) for $h_1=h_2=h_3$.

The exponentially growing solution illustrates the Ostrogradsky instability. As the evolution equations are reversible, the exponentially growing solution is accompanied by an exponentially decaying solution. The growth and decay rates are determined by the time scale $\lambda$, which is not in the original problem and enters the picture only through the transformation (\ref{exlintransf}). One could avoid the instability by taking the limit $\lambda \rightarrow \infty$, in which $\bar{q}_1$, $\bar{q}_2$ are only shifted by constants. As these constants are unphysical, it seems preferable to avoid exponentially growing solutions by imposing suitable constraints, as proposed in the general development.

For the Hamiltonian formulation of the problem, we rely on the correspondences implied by Eqs.~(\ref{ostrovarQ12}), (\ref{qbarddot}) and (\ref{qbardddot}),
\begin{equation}\label{exLHcor1a}
   Q_{11} = \bar{q}_1 , \quad
   Q_{21} = \dot{\bar{q}}_1 , \quad
   P_{21} = \lambda^2 m \ddot{\bar{q}}_1 ,
\end{equation}
\begin{equation}\label{exLHcor1b}
   P_{11} = m ( \dot{\bar{q}}_1 + \lambda \ddot{\bar{q}}_2 - \lambda^2 \qbarddd_1 )
   - \lambda^2 h \dot{\bar{q}}_1 ,
\end{equation}
\begin{equation}\label{exLHcor2a}
   Q_{12} = \bar{q}_2 , \quad
   Q_{22} = \dot{\bar{q}}_2 , \quad
   P_{22} = \lambda m ( \dot{\bar{q}}_1 + \lambda \ddot{\bar{q}}_2 ) ,
\end{equation}
\begin{equation}\label{exLHcor2b}
   P_{12} = m ( \dot{\bar{q}}_2 - \lambda \ddot{\bar{q}}_1 - \lambda^2 \qbarddd_2 )
   - \lambda h ( \bar{q}_1 + \lambda \dot{\bar{q}}_2 ) .
\end{equation}
The Hamiltonian is found to be of the quadratic form
\begin{eqnarray}
   H &=& \frac{1}{2 \lambda^2 m} (P_{21}^2 + P_{22}^2)
   + \left(P_{11}-\frac{P_{22}}{\lambda}\right) Q_{21} + P_{12} Q_{22}
   \nonumber \\
   &-& \frac{1}{2} m Q_{22}^2
   + \frac{1}{2} h [ (Q_{11}+\lambda Q_{22})^2 + Q_{12}^2 + \lambda^2 Q_{21}^2 ] ,
   \nonumber \\ &&
\label{exH}
\end{eqnarray}
where the last term (proportional to the parameter $h$) is the potential $\bar{V}$. Most of the evolution equations resulting from this Hamiltonian in a canonical way can be interpreted as assignments of variables,
\begin{equation}\label{exHamevolas1}
  \dot{Q}_{11} = Q_{21} , \qquad
  \dot{Q}_{12} = Q_{22} , ,
\end{equation}
\begin{equation}\label{exHamevolas2}
  \dot{Q}_{21} = \frac{P_{21}}{\lambda^2 m} , \qquad
  \dot{Q}_{22} = \frac{P_{22}}{\lambda^2 m} - \frac{Q_{21}}{\lambda} ,
\end{equation}
and
\begin{eqnarray}
  \dot{P}_{21} &=& \frac{P_{22}}{\lambda} - P_{11} - \lambda^2 h Q_{21} ,
  \nonumber \\
  \dot{P}_{22} &=& m Q_{22} - \lambda h (Q_{11}+\lambda Q_{22}) - P_{12} . \qquad
\label{exHamevolas3}
\end{eqnarray}
The final two evolution equations
\begin{equation}\label{exHamevol}
  \dot{P}_{11} = - h (Q_{11}+\lambda Q_{22}) , \qquad
  \dot{P}_{12} = - h Q_{12} ,
\end{equation}
coincide with the two fourth-order differential equations of the Lagrangian approach given in Eq.~(\ref{exLageq4}).

If we impose the four constraints
\begin{eqnarray}
   \varphi_1 &=& \lambda P_{11} - P_{22} = 0 ,
   \nonumber \\
   \varphi_2 &=& \lambda \dot{\varphi}_1 = \lambda ( P_{12} - m Q_{22} ) = 0 ,
   \nonumber \\
   \varphi_3 &=& \lambda \dot{\varphi}_2 = - P_{22} + \lambda m Q_{21} - \lambda^2 h Q_{12} = 0 ,
   \nonumber \\
   \varphi_4 &=& \lambda \dot{\varphi}_3 = \varphi_2 + P_{21} + \lambda^2 h Q_{11} = 0 ,
\label{exconstr}
\end{eqnarray}
the identity
\begin{equation}\label{exconstrc}
   \lambda \dot{\varphi}_4 = \varphi_3 - \varphi_1 ,
\end{equation}
implies that the Poisson bracket of the Hamiltonian with each of these constraints vanishes on the constrained manifold. The first two constraints in Eq.~(\ref{exconstr}) actually coincide with the primary constraints (\ref{genconstraints1}) of the general development (except for a factor of $\lambda$). The second constraint arises also among the secondary constraints (\ref{genconstraints2}), so that this classification scheme clearly comes with ambiguities. In the general development, the fourth constraint in Eq.~(\ref{exconstr}) arises already as a tertiary constraint, after which the iterative procedure comes to an end.

By means of the first two constraints, which are the primary constraints of the general development, the Hamiltonian (\ref{exH}) on the constrained manifold can be written as
\begin{eqnarray}
   H &=& \frac{1}{2 \lambda^2 m} (P_{21}^2 + P_{22}^2) + \frac{1}{2} m Q_{22}^2
   \nonumber \\
   &+&  \frac{1}{2} h [ (Q_{11}+\lambda Q_{22})^2 + Q_{12}^2 + \lambda^2 Q_{21}^2 ] , \qquad
\label{exHconstr}
\end{eqnarray}
which (for $m,h > 0$) is clearly bounded from below by zero. Indeed, the four constraints eliminate all the exponentially growing or decaying terms from the solutions (\ref{solqbarigen1}) and (\ref{solqbarigen2}) ($\bar{c}_5=\bar{c}_6=\bar{c}_7=\bar{c}_8=0$), so that only the two harmonic oscillators (\ref{solqbari}) resulting from the original system of three harmonic oscillators survive in the higher-order theory. The reduction from three to two harmonic oscillators corresponds to the reduction from $K$ to $I$ basic variables in the higher-order theory, where the reduction from three to two plays a special role because it allows for a one-to-one correspondence of the full sets of variables in the Lagrangian, as pointed out in the paragraph after Eq.~(\ref{albeder}).

The matrix of Poisson brackets among the constraints,
\begin{equation}\label{PBconstr}
   (\Pbracket{\varphi_k}{\varphi_l}) = \lambda m
   \left(
     \begin{array}{cccc}
       0 & -1 & 0 & -1-\Omega \\
       1 & 0 & 1+\Omega & 0 \\
       0 & -1-\Omega & 0 & - \Omega \\
       1+\Omega & 0 & \Omega & 0 \\
     \end{array}
   \right) ,
\end{equation}
where $\Omega=\lambda^2 \omega^2$, implies that we deal with second class constraints, for which a Dirac bracket can be introduced and a canonical quantization procedure is available \cite{Dirac50,Dirac58a,Dirac58b}. As expected from general grounds, the matrix in Eq.~(\ref{PBconstr}) is invertible for all $\Omega$. Only two $2 \times 2$ submatrices need to be inverted, so hat the Dirac bracket can be written down in closed form.

\section{Example from field theory} \label{appexampleGR}
We here look at the weak-field approximation for a Yang-Mills theory based on the Lorentz group, which has been proposed as an alternative theory for gravity \cite{hco231}. For this field theoretic example, the transformation of the type (\ref{variabletransf}) is given by
\begin{equation}\label{exGRAdef}
   A_{\kappa\lambda\nu} = \frac{1}{2} \left(
   \frac{\partial h_{\lambda\nu}}{\partial x^\kappa}
   - \frac{\partial h_{\kappa\nu}}{\partial x^\lambda} \right) ,
\end{equation}
where the Greek letters denote space-time indices (with the standard convention $x^0=ct$, where $c$ is the speed of light). The space and time dependent field $h_{\lambda\nu}$ can be interpreted as the deviation of the metric from the Minkowski metric [with signature $(-,+,+,+)$, which we use for lowering or raising indices]. Note that $A_{\kappa\lambda\nu}$ is antisymmetric in $\kappa$ and $\lambda$ so that only six index combinations matter. These six index pairs correspond to the generators of the Lorentz group, where the pairs $(0,1)$, $(0,2)$, $(0,3)$ correspond to boosts in the three coordinate directions and $(2,3)$, $(3,1)$, $(1,2)$ correspond to rotations around the coordinate axes, indicated by the corresponding perpendicular planes. Therefore, in the definition of $A_{\kappa\lambda\nu}$, the pairs $(\kappa,\lambda)$ label the generators, and $\nu$ is interpreted as the space-time index of the vector potential from which a Yang-Mills field tensor can be defined.

To keep the example simple, we here assume that only the spatial components,
\begin{equation}\label{exGRqbardef}
   \bar{q}_{kl} = h_{kl} ,
\end{equation}
are needed to parametrize the vector potential, that is, we choose $h_{00}=h_{0k}=h_{k0}=0$ in Eq.~(\ref{exGRAdef}). According to Section~10.2 of \cite{Weinberg}, this setting is sufficient for discussing gravitational waves. All fields $A_{\kappa\lambda\nu}$ then vanish for $\nu=0$. The vector fields associated with rotations are linear in $h_{kl}$ [including spatial derivatives, which belong to the $\alpha$ part of the transformation (\ref{variabletransf})], whereas the vector fields associated with boosts are linear in the time derivative of $h_{kl}$ [and therefore belong to the $\beta$ part of the transformation (\ref{variabletransf})]. For symmetric $h_{kl}$, there are six independent basic variables.
In the weak-field approximation, the Yang-Mills Lagrangian associated with the vector potentials (\ref{exGRAdef}) is given by the space integral of the following quadratic Lagrangian density,
\begin{eqnarray}
   {\cal L} &=&
   \frac{1}{8}
   \frac{\partial^2  h^{\mu\nu}}{\partial x_\kappa \partial x_\lambda}
   \nonumber \\ & \times & \left(
   \frac{\partial^2  h_{\mu\nu}}{\partial x^\kappa \partial x^\lambda} +
   \frac{\partial^2  h_{\kappa\lambda}}{\partial x^\mu \partial x^\nu} -
   \frac{\partial^2  h_{\kappa\nu}}{\partial x^\mu \partial x^\lambda} -
   \frac{\partial^2  h_{\mu\lambda}}{\partial x^\kappa \partial x^\nu} \right) .
   \nonumber \\ &&
\label{exGRLexp}
\end{eqnarray}
One half of the expression in parentheses is equal to the Yang-Mills field tensor. The Lagrangian density (\ref{exGRLexp}) leads to the fourth-order field equation
\begin{equation}\label{exGRfieldeq}
   \Box\Box  h_{kl} -
   \Box\frac{\partial^2  h_{nl}}{\partial x^k \partial x_n} -
   \Box \frac{\partial^2  h_{kn}}{\partial x_n \partial x^l} +
   \frac{\partial^4  h_{mn}}{\partial x_m \partial x_n \partial x^k \partial x^l} = 0 ,
\end{equation}
where $\Box = \partial^2 / \partial x^\mu \partial x_\mu$. This field equation can be rewritten in the elegant alternative form
\begin{equation}\label{weakfieldR}
   \Box R_{kl}^{(1)} - \frac{1}{2}
   \frac{\partial^2 R^{(1)}}{\partial x^k \partial x^l} = 0 ,
\end{equation}
where $R_{\mu\nu}^{(1)}$ and $R^{(1)}$ are the linearized versions of the Ricci tensor associated with $h_{\mu\nu}$ and the corresponding curvature scalar.

For the Hamiltonian formulation, we find the canonical variables
\begin{equation}\label{exGRQdef}
   Q_{1kl} = h_{kl} , \quad
   Q_{2kl} = \frac{\partial h_{kl}}{\partial t},
\end{equation}
\begin{equation}\label{exGRP2def}
   P_{2kl} = \frac{1}{4c^4} \frac{\partial^2  h_{kl}}{\partial t^2} ,
\end{equation}
and
\begin{equation}\label{exGRP1def}
   P_{1kl} = \frac{1}{4c^2} \frac{\partial}{\partial t}
   \left( \Box  h_{kl} +
   \frac{\partial^2  h_{kl}}{\partial x_n \partial x^n} -
   \frac{\partial^2  h_{nl}}{\partial x^k \partial x_n} -
   \frac{\partial^2  h_{kn}}{\partial x_n \partial x^l} \right) ,
\end{equation}
which, through the time derivative of $\Box  h_{kl}$, includes the third-order time derivative of $h_{kl}$. The Hamiltonian density is found to be
\begin{eqnarray}
   {\cal H} &=&
   2 c^4 P_{2kl} P_{2kl} + P_{1kl} Q_{2kl}
   - \frac{1}{8} \frac{\partial^2 Q_{1kl}}{\partial x^m \partial x^n}
   \nonumber \\
   & \times &
   \bigg( \frac{\partial^2 Q_{1kl}}{\partial x^m \partial x^n}
   +  \frac{\partial^2 Q_{1mn}}{\partial x^k \partial x^l}
   -  \frac{\partial^2 Q_{1ml}}{\partial x^k \partial x^n}
   - \frac{\partial^2 Q_{1kn}}{\partial x^m \partial x^l}
   \bigg)
   \nonumber \\
   &+& \frac{1}{8c^2} \frac{\partial Q_{2kl}}{\partial x^n}
   \left( 2 \frac{\partial Q_{2kl}}{\partial x^n}
   - \frac{\partial Q_{2nl}}{\partial x^k}
   - \frac{\partial Q_{2kn}}{\partial x^l} \right) .
\label{exGRHexp}
\end{eqnarray}
Three of the canonical Hamiltonian equations reproduce the time derivatives of the quantities in Eqs.~(\ref{exGRQdef}) and (\ref{exGRP2def}). The Hamiltonian evolution equation for $P_{1kl}$ reproduces the field equation (\ref{exGRfieldeq}).

In order to find the primary constraints for eliminating the Ostrogradsky instability we need to go back to the general development and the corresponding structure of Eq.~(\ref{genconstraints1}). For our field theoretic example, these primary constraints become
\begin{equation}\label{exGRprimconstr}
   P_{1kl} - \frac{1}{8c^2} \frac{\partial}{\partial x_n}
   \left( 2 \frac{\partial Q_{2kl}}{\partial x^n}
   - \frac{\partial Q_{2nl}}{\partial x^k}
   - \frac{\partial Q_{2kn}}{\partial x^l} \right) = 0 ,
\end{equation}
or, with the help of Eqs.~(\ref{exGRQdef}) and (\ref{exGRP1def}),
\begin{equation}\label{exGRprimconstrh}
   \frac{\partial}{\partial t}
   \left( \Box  h_{kl} -
   \frac{1}{2} \frac{\partial^2  h_{nl}}{\partial x^k \partial x_n} -
   \frac{1}{2} \frac{\partial^2  h_{kn}}{\partial x_n \partial x^l} \right) = 0 .
\end{equation}
The secondary constraints obtained as the time derivative of Eq.~(\ref{exGRprimconstrh}) can be used to eliminate the fourth-order time derivatives from the field equation (\ref{exGRfieldeq}) to find a second-order differential equation in time.

The possibility of identifying the primary constraints is the key advantage of the class of higher-order models introduced in this paper. We illustrate their importance for the special case of gravitational waves. If we assume transverse waves, the field equation (\ref{exGRfieldeq}) is reduced to
\begin{equation}\label{exGRfieldeqwaves}
   \Box\Box  h_{kl} = 0 ,
\end{equation}
and the primary constraints (\ref{exGRprimconstrh}) become
\begin{equation}\label{exGRprimconstrhwaves}
   \frac{\partial}{\partial t} \Box  h_{kl} = 0 .
\end{equation}
Both equations are satisfied by the plane wave solutions characterized by $\Box  h_{kl} = 0$, but additional instable solutions need to be excluded from the higher-order equations. The combination of Eqs.~(\ref{exGRfieldeqwaves}) and (\ref{exGRprimconstrhwaves}) gives $\Delta \Box  h_{kl} = 0$, where $\Delta$ is the Laplacian. With suitable spatial boundary conditions one could arrive at the desired equation $\Box  h_{kl} = 0$.

Instead of expressing the vector potentials in terms of the spatial components $h_{kl}$ it would be preferable to use the full deviatoric metric four-tensor $h_{\mu\nu}$ to keep the Lorentz covariance in the description. However, such an approach would be significantly more challenging because an explicit consideration of gauge invariance and the associated constraints would be required. These extra efforts should certainly be made if one wishes to go beyond the weak-field approximation used here for illustrative purposes.


\vfill


%

\end{document}